\newcommand{\phibar}{\bar{\phi}}
\newcommand{\phibardot}{\dot{\bar{\phi}}}

\newcommand{\gtilde}{\tilde{g}_{\mu \nu}}
\newcommand{\gtildecon}{\tilde{g}^{\mu \nu}}

\newcommand{\beq}{\begin{equation}}
\newcommand{\eeq}{\end{equation}}
\newcommand{\beqarr}{\begin{eqnarray}}
\newcommand{\eeqarr}{\end{eqnarray}}
\newcommand{\ec}[1]{Eq.~(\ref{eq:#1})}

\newcommand{\eql}[1]{\label{eq:#1}}
\def\bea{\begin{eqnarray}}
\def\eea{\end{eqnarray}}
\newcommand{\vs}{\nonumber\\}
\def\be{\begin{equation}}
\def\ee{\end{equation}}

\documentclass[aps,nofootinbib,twocolumn,prl]{revtex4} 
\usepackage{graphicx}
\newcommand{\sfig}[2]{
\includegraphics[width=#2]{#1}
        }
\newcommand{\Sfig}[2]{
    \begin{figure}[thbp]
    \sfig{#1.eps}{0.9\columnwidth}
    \caption{{\small #2}}
    \label{fig:#1}
    \end{figure}
}
\newcommand{\Rf}[1]{\ref{fig:#1}}
\newcommand{\rf}[1]{\ref{fig:#1}}

\begin{document}

\title{Can Cosmic Structure form without Dark Matter?}

\author{Scott Dodelson$^{1,2}$ and Michele Liguori$^{1,3}$}

\affiliation{$^1$Particle Astrophysics Center,
Fermi National Accelerator Laboratory, Batavia, IL~~60510-0500}
\affiliation{$^2$Department of Astronomy \& Astrophysics, The University of Chicago,
Chicago, IL~~60637-1433}
\affiliation{$^3$Dipartimento di Fisica "G. Galilei" Universit\`a di Padova,
INFN sezione di Padova, Via Marzolo 8, I-35131 Padova, Italy}

\date{\today}
\begin{abstract}
One of the prime pieces of evidence for dark matter is the observation
of large overdense regions in the universe. Since we know from the cosmic
microwave background that the regions that contained the most baryons
when the universe was $\sim400,000$ years old were overdense
by only one part in ten thousand, perturbations had to have grown since then
by a factor greater than $(1+z_*)\simeq 1180$ where $z_*$ is the epoch of 
recombination. This enhanced growth does not happen in general relativity, 
so dark matter is
needed in the standard theory. We show here that 
enhanced growth can occur in alternatives to general relativity, 
in particular in
Bekenstein's relativistic version of MOdified Newtonian Dynamics (MOND).
The vector field introduced in that theory for a completely different reason 
plays a key role in generating the instability that produces large cosmic
structures today.
\end{abstract}

\maketitle

{\parindent0pt\it Introduction.}
Dark Matter was introduced long ago to explain galactic
rotation curves~\cite{1978ApJ...225L.107R} and large velocities 
in galaxy clusters~\cite{1937ApJ....86..217Z}. Over the past decade
the case for dark matter has gotten stronger: galactic rotation curves still
appear to diverge significantly from what is expected from observations of
visible matter; gravitational lensing is able to map the mass distribution in a
galaxy or cluster and this mass distribution often does not coincide with the
luminous matter~\cite{Clowe:2006eq}; and large overdense regions in the universe can be
explained~\cite{1985ApJ...292..371D} only if dark matter was around early on to seed structure formation
when the universe was of order several hundred thousand years old. At the same time,
the case that dark matter does {\it not} consist simply of protons and neutrons
which do not emit light has also gotten stronger~\cite{Kirkman:2003uv,Steigman:2003ey}. 
To explain the astronomical and cosmological
observations, therefore, we apparently need to introduce a new fundamental particle which
has not yet been observed in accelerators.

There is one way of avoiding this conclusion: perhaps the implicit assumption that gravity
is described by general relativity is incorrect. Perhaps a fundamental theory of gravity
which differs from general relativity on large scales can explain the observations without
recourse to new, unobserved particles. Now more than ever before, there are very good reasons
to explore this idea of modifying gravity. For, the case for dark energy also hinges on the
assumption that general relativity describes gravity on large scales. Dark energy is even more 
difficult to explain in the context of fundamental theories than is dark matter, so it seems almost
natural to look at gravity as the culprit in both cases. 

Perhaps the most direct piece of evidence for dark matter is from galactic rotation
curves, which are much flatter than the $R^{-1/2}$ fall-off expected from
observations of visible matter. Over 25 years ago, Milgrom~\cite{Milgrom:1983pn} proposed a MOdified Newtonian
Dynamics (MOND), which diverges from Newtonian theory when the gravitational acceleration
is less than $a_0\sim (200 {\rm km\,sec}^{-1})^2/(10\,{\rm kpc})$. Argument has raged
for two decades as to whether this modification 
is consistent with a wide variety of observations~\cite{Sanders:2002pf,Clowe:2003tk,Moffat:2004bm,Sellwood:2004xu,Pointecouteau:2005mr}.
Part of the difficulty in assessing MOND is that it makes no claims to be a comprehensive theory
of gravity, so, for example, it is impossible to test it with cosmological observations.

Recently, Bekenstein~\cite{Bekenstein} constructed a fully relativistic covariant theory which reduces to MOND in the
appropriate (static) limit. In addition to the gravitational metric (a tensor field), Bekenstein's theory contains
a scalar field and a vector field; hence he called it TensorVectorScalar, or TeVeS. 
We can now compare the predictions of TeVeS to
those of general relativity. There are two features of the new theory that are particularly 
interesting and seem worthy of further
study. First, several new constants must be introduced into the TeVeS Lagrangian and one of these 
is of order $a_0$, the MOND
scale. This is no surprise of course since TeVeS is designed to reduce to MOND in certain limits. 
The interesting part is that $a_0$
is roughly the same order of magnitude as the fundamental scale introduced in quintessence or modified gravity 
theories designed to explain
the acceleration of the universe. The motivation for TeVeS had nothing to do with the cosmic acceleration. 
Is it just a coincidence then
that the fundamental scale needed is the one required to obtain acceleration? 
Or is this an indication that we are on the right track
in our quest to explain away the dark sector by modifying gravity? 

\Sfig{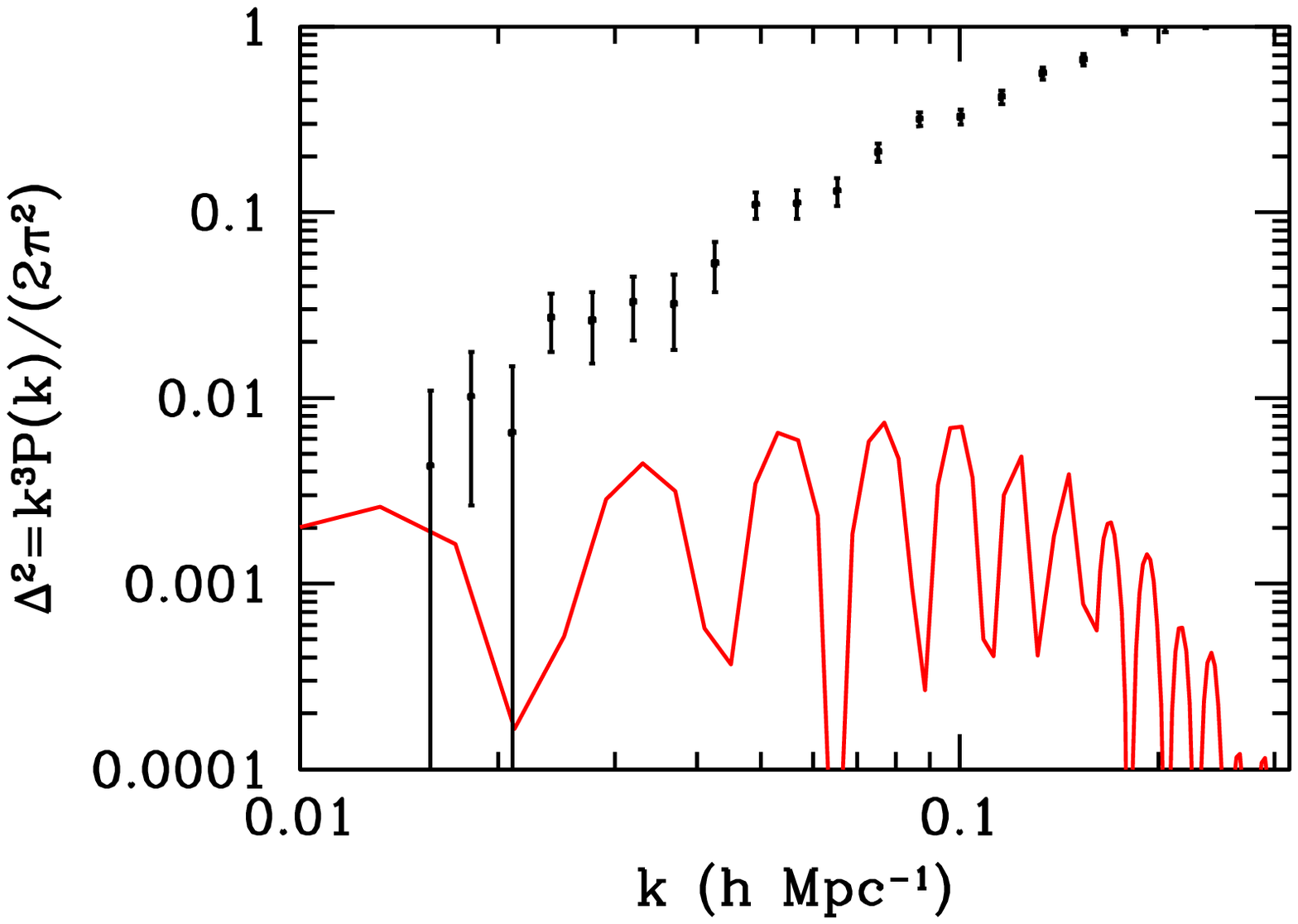}{Power spectrum of matter fluctuations in a theory without dark matter as compared to observations of the galaxy
power spectrum. The observed spectrum~\cite{Tegmark:2003uf} does not have the pronounced wiggles predicted by a baryon-only model, but it also has significantly
higher power than does the model. In fact $\Delta^2$, which is a dimensionless measure of the clumping, never rises above one in 
a baryon-only model, so we would not expect to see any large structures (clusters, galaxies, people, etc.) in the universe in such a model.}

The second intriguing aspect of TeVeS, and the one we focus on here, 
was recently uncovered by Skordis and
collaborators~\cite{Skordis}. To understand the importance of this feature, it is necessary to 
state an obvious hurdle
that any no-dark-matter theory must overcome. This hurdle is best depicted in 
Figure~\rf{powernocdm} which shows the observed power spectrum
of matter in the universe in a theory without non-baryonic dark matter. 
This theory fails to describe observations in two ways: (i) the
shape of the power spectrum is off and (ii) the amplitude is far too small. The first failure 
has been exploited by many authors
to prove the existence of non-baryonic dark matter
~\cite{Tegmark:2003ud,Spergel:2006hy}, 
the statistical significance for which now exceeds 5-sigma. The second
failure is often ignored because analysts typically marginalize over the amplitude of the power 
spectrum on the grounds that
the power spectrum of galaxies is likely to differ by an overall normalization factor (the bias) from 
the power spectrum of matter.
But a baryon-only model fails miserably at getting anywhere near the amplitude required to generate 
galaxies and galaxy clusters even with
an absurd amount of bias. So if we really want to do away with dark matter, 
we need to find a mechanism of growing perturbations faster
than in standard general relativity. This is precisely what Skordis et al.~\cite{Skordis,Skordis2} seemed 
to have found in their 
treatment of perturbations around a smooth cosmological solution in TeVeS. Here we aim to move beyond 
their numerical treatment to isolate
what is causing enhanced growth. Our motivation goes beyond TeVeS, as the 
exact Lagrangian in \cite{Bekenstein} will almost
certainly need to be altered even if the general idea turns out to be correct. Indeed, as shown in Fig~\rf{powernocdm}, 
even
if structure grows faster than in the standard theory, 
the shape of the baryon-only spectrum does not match the observations.
Rather, we want to understand generally how to modify
gravity such that it solves not only the galactic rotation curve problem but also the cosmological structure problem.

{\parindent0pt\it Cosmology in TeVeS.}
Ordinary matter couples to the gravitational metric $g_{\mu\nu}$ in the standard way in the TeVeS model. 
The metric which couples to matter, though, does not appear in the standard way in the Einstein-Hilbert
action. Rather, it is useful to define a new tensor $\tilde{g}_{\mu \nu}$ which is a functional of
 $g_{\mu\nu}$ and a scalar field $\phi$ and a vector field $A_\mu$. Specifically, 
\beq\label{eqn:g2gtilde}
g_{\mu \nu} \equiv e^{-2 \phi} \left(\tilde{g}_{\mu \nu} +
A_\mu A_\nu \right)- e^{2 \phi} A_\mu A_\nu \, 
\eeq
defines $\tilde g_{\mu\nu}$. The action of $\gtilde$ is the standard Einstein-Hilbert action.
The scalar and vector fields have dynamics given, 
respectively, by the actions $S_s$ and $S_v$:
\begin{eqnarray}
S_s & = & \frac{-1}{16 \pi G}  \int d^4 x (-\tilde{g})^{1/2}
          \left[\mu \left(\gtildecon - A^\mu A^\nu \right)
          \phi_{,\mu} \phi_{\nu} + V\right]  \vs
S_v & = & \frac{-1}{32 \pi G} \int d^4 x (-\tilde{g})^{1/2}
          \left[K F^{\alpha \beta} F_{\alpha \beta} 
	    - 2 \lambda \left(A^2 + 1 \right) \right] 
\end{eqnarray}
where $\mu$ is an additional non-dynamical 
scalar field, 
$F_{\mu \nu} \equiv A_{\mu,\nu} - A_{\nu,\mu}$, and indices are raised and lowered with 
the metric $\gtilde$. The potential 
$V(\mu)$ is chosen to give the correct non-relativistic
MONDian limit. We will consider the form proposed 
by Bekenstein~\cite{Bekenstein}:
\beq
V = \frac{3 \mu_0^2}{128 \pi \, \ell_B^2}
    \left[\hat{\mu} (4 + 2 \hat{\mu} - 4 \hat{\mu} + \hat{\mu}^3 )
          + 2 \ln{(\hat{\mu}-1)^2} \right] \;
\eeq
with $\hat\mu\equiv \mu/\mu_0$. There are three free parameters that appear in 
the TeVeS action: $\mu_0$, 
$\ell_B$ and $K_B$. The parameter $\lambda$ in the vector field action is completely fixed 
by variation of the action.

Armed with this action, we can solve~\cite{Bekenstein,Skordis} for the evolution of the scale factor $a$ of a homogeneous
Friedman-Robertson-Walker (FRW) metric. This evolution turns out to be very similar to the
standard case, with several small deviations. First, Newton's constant gets generalized to
$Ge^{-4\phi}/(1+d\phi/d\ln(a))^2$. Second, the Friedman equation governing the evolution of $a$ has, in addition to
the standard source terms of the matter and radiation energy densities, the energy density of $\phi$: 
\beq
\rho_\phi={e^{2\phi}\over 16\pi G} \left( \mu V' + V\right)
.\eql{rhophi}\eeq

\Sfig{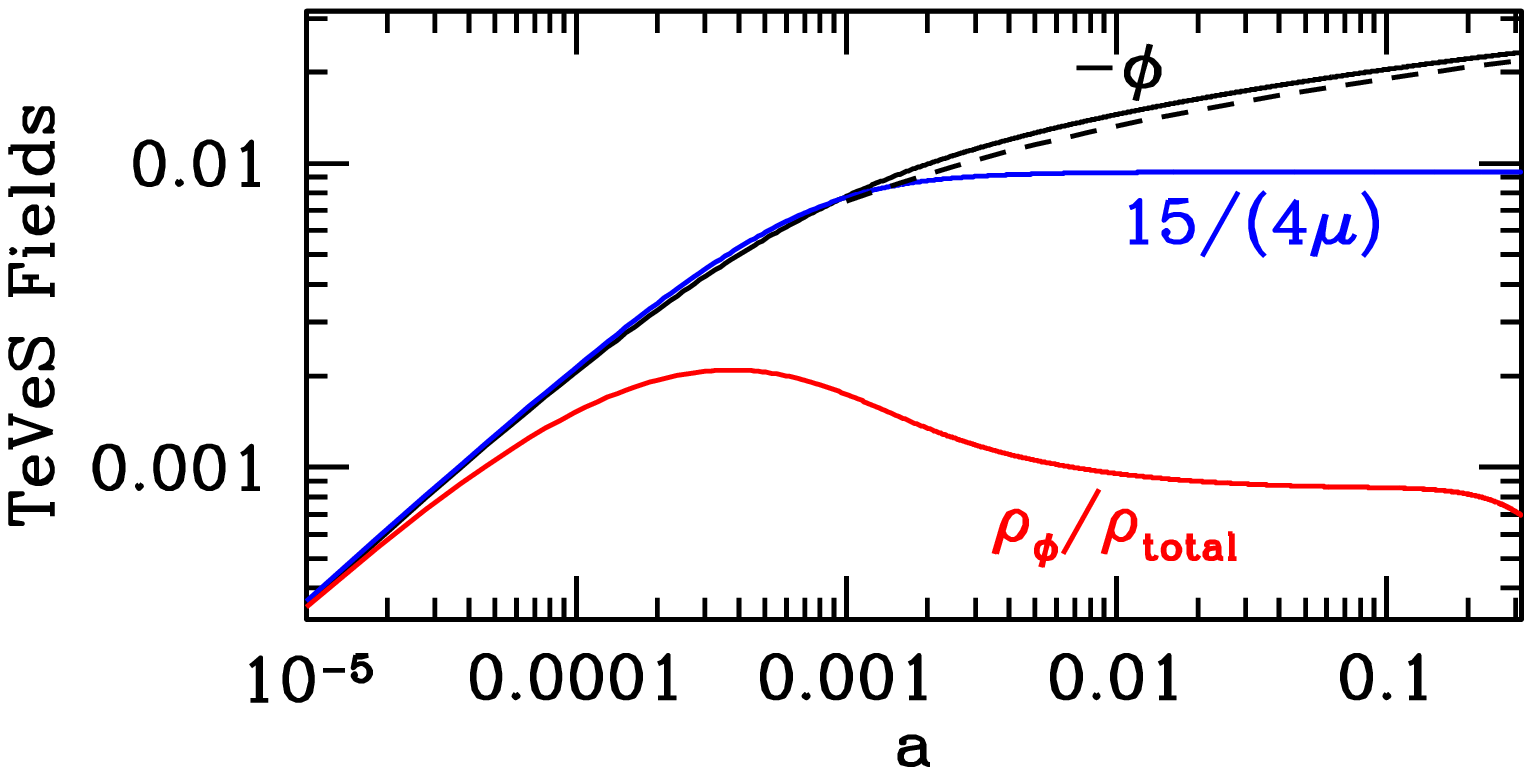}{Evolution of homogeneous TeVeS fields. Dashed line shows logarithmic approximation for
$\phi$ valid in the regime when $\mu$ is constant. In that regime, $\rho_\phi$ scales as the ambient
density, with the ratio equal to $(6\mu_0)^{-1}$ in the matter era. Early on, $\rho_\phi/\rho_{\rm
total}=-\phi=15/(4\mu)$.}

The TeVeS modifications to the standard cosmology then depend on the evolution of the scalar field $\phi$. 
During the
radiation dominated era, $\rho_\phi$ is much smaller than the dominant radiation density, 
with the ratio growing as
$a^{4/5}$~\cite{us}. As shown in Fig.~\rf{rhophi}, once $a$ gets large enough, $\mu$ becomes constant and
$\phi\simeq {\rm constant}\,+\ln(a)/2\mu_0$. The energy density in $\phi$ 
then scales as $a^{-3}$ exactly like matter, with the
ratio fixed at $(6\mu_0)^{-1}$. For a variety of reasons~\cite{Bekenstein}, 
large values of $\mu_0 (\sim 100-1000)$ are
preferred, so $\rho_\phi$ is much smaller than the
other densities at all times and the effective Newton 
constant differs at only the percent level from the standard one.
Homogeneous expansion is there the same in TeVeS as in standard general relativity.

{\parindent0pt\it Perturbations in TeVeS.}
We start the perturbation expansion by writing the metric which couples to matter as
\begin{eqnarray}
g_{00}(\vec x,\tau) & = & -a^2(\tau)(1 - 2 \Psi(\vec x,\tau))  \\
g_{ij}(\vec x,\tau) & = & a^2(\tau)(1+2\Phi(\vec x,\tau))\delta_{ij} 
\end{eqnarray}
where $\tau$ is conformal time; similarly we
expand the density of the matter (and/or radiation) field as 
\beq
\rho(\vec x,\tau)  = \bar{\rho}(\tau)\left(1 + \delta(\vec x,\tau)\right) 
.\eeq

To consider the evolution of perturbations in TeVeS, we need to perturb not only the matter
and the FRW metric fields, but also the new fields: $\phi$ and $A_\mu$~\cite{Skordis,Skordis2}.
The scalar perturbation can be written as
\beq
\phi(\vec x,\tau)  =  \phibar(\tau) + \varphi(\vec x,\tau)
\eeq
where $\phibar$ is the zero order field introduced above. 

The vector field requires a little more thought~\cite{Lim:2004js,Zlosnik:2006zu}.
In general, perturbations to a vector field $A^\mu$ will be described by four independent functions. In
this case, though, the two transverse spatial components decouple from the set of scalar perturbation
equations, so we can neglect them. Further simplification follows from the fact
that the vector field in TeVeS is subject to the constraint $A^\mu A_\mu\equiv 
\tilde g^{\mu\nu}A_\mu A_\nu = -1$. This fixes the time component of the perturbation, so we need track
only the longitudinal component of $A^\mu$. In detail, 
the zero order $A_\mu$ can be chosen to have only a time component. The constraint then 
sets that time component
to $a e^{-\phibar}$ (since $\tilde g^{00}=-a^{-2} e^{2\phibar}$), so the perturbed vector field
can be written
\beq
A_\mu(\vec x,\tau)   =  a(\tau) e^{-\phibar(\tau)} \left(\bar{A}_\mu + \alpha_\mu\right) 
\eeq
where $\bar{A}_{\mu} \equiv (1,0,0,0)$ and
\beq
\alpha_{\mu} \equiv \left(\Psi + \varphi, \vec{\alpha}\right)
.\eeq 
That is, the time component of the perturbation is constrained to be a combination of the perturbations
to the metric ($\Psi$) and the scalar field ($\varphi$). 
Skorids {\it et al.}\,\cite{Skordis} called the longitudinal component of the 
perturbation $\alpha$ (with no
index); specifically, $\vec\nabla \alpha \equiv \vec\alpha$ or equivalently
$\nabla^2\alpha \equiv \vec\nabla\cdot \vec\alpha$. 

The perturbations to the metric, matter, radiation, and TeVeS fields are governed by
a set of coupled differential equations. Ref.~\cite{Skordis} suggested that perturbations in the
scalar field may induce enhanced growth in the matter perturbations. We have found~\cite{us} that this is not so:
the scalar perturbations are small early on and then oscillate about $\Phi/\mu_0$ in the matter epoch.
This is far too small a value to impact density perturbations.

Rather we find~\cite{us} that the vector perturbations are the key to enhanced growth.
The equation governing
vector perturbations is:
\beq
\ddot\alpha + b_1\dot\alpha + b_2\alpha=S[\Phi,\Psi]
\eql{ve}
\eeq
where the source term on the right is a functional of the metric perturbations and
the coefficients on the left are
\bea
b_1&\equiv&{\dot a\over a} + 5\phibardot - a_1 \vs
b_2&\equiv&-a_1\left( 5\phibardot+{\dot a\over a} \right) -\dot a_1
\vs
&& +e^{-4\phi} K_B^{-1} \left[ 8\pi G(\rho+P)a^2(1-e^{-4\phibar}) -\mu\dot\phi^2 \right]
.\eql{defb}
\eea
Here $a_1$ is defined as
\be
a_1\equiv \phibardot\left(e^{-4\phibar} - 4 \right)
+{\dot a\over a} \left(e^{-4\phibar} - 2 \right)
.\ee

There is one limit in which analytic solutions exist for the homogeneous part of \ec{ve}: when the background
quantities are such that $\phibar$ and $\phibardot$ vanish. This is a fairly good
approximation because we know that $\phi$ has little impact on the zero order
dynamics. In this limit, $b_1\rightarrow 2\dot a/a$ and $b_2\rightarrow \ddot
a/a$. For simplicity then consider a matter dominated universe so that
$b_1=4/\tau$ and $b_2=2/\tau^2$. For reasons that will become clear soon, let
us write
\be
b_1={4\over \tau} \qquad b_2={2(1+\epsilon)\over \tau^2}
.\ee
In the limit we are now considering, $\epsilon=0$. 

The homogeneous solutions to \ec{ve} scale as $\tau^p$ with the powers determined by
solving an algebraic equation so that
\be
p_{\pm} ={-3\over 2} \pm {1\over 2} \sqrt{9-8(1+\epsilon)}
.\eql{power}\ee
So when $\epsilon=0$, the two homogeneous modes scale as $\tau^{-2}$ and $\tau^{-1}$. The particular solution
will then dominate; in the matter era, the dominant source term is $-6\Psi/\tau$, so $\alpha=-\Psi\tau/3$.
The top panel of Fig.~\Rf{allpert} shows that, when $K_B$ is not too small, $\alpha$ does indeed follow
the particular solution.

However, when $K_B$ is small, the term in \ec{defb} multiplied by $K_B^{-1}$ cannot
be neglected. The ratio of this term to $\ddot a/a$ is defined as
$\epsilon$. Both terms in square brackets scale the same way,
but the first term is quite a bit larger. In a matter dominated universe,
$8\pi G(\rho+P) a^2 = 3a^2H^2 \rightarrow 12/\tau^2$, so using the logarithmic solution
for $\phibar$, we find
\be
\epsilon \simeq 
{-12 \ln\left( a/5\times10^{-5} \right) \over \mu_0K_B} 
.\eql{eps}\ee
So as long as $K_B\mu_0$ is of order 100 or smaller, the coefficient of $\alpha$ in \ec{ve}, $1+\epsilon$, will eventually 
become negative, signifying enhanced growth. Note that because of the slow, logarithmic growth of $\phibar$,
the enhanced growth will not be exponential. It will, however, be a growing mode, which deviates from the particular solution
$-\Psi\tau/3$. The top panel of Figure~\Rf{allpert} shows that the qualitative aspects of this analytic solution do emerge in
the full numerical results.

\Sfig{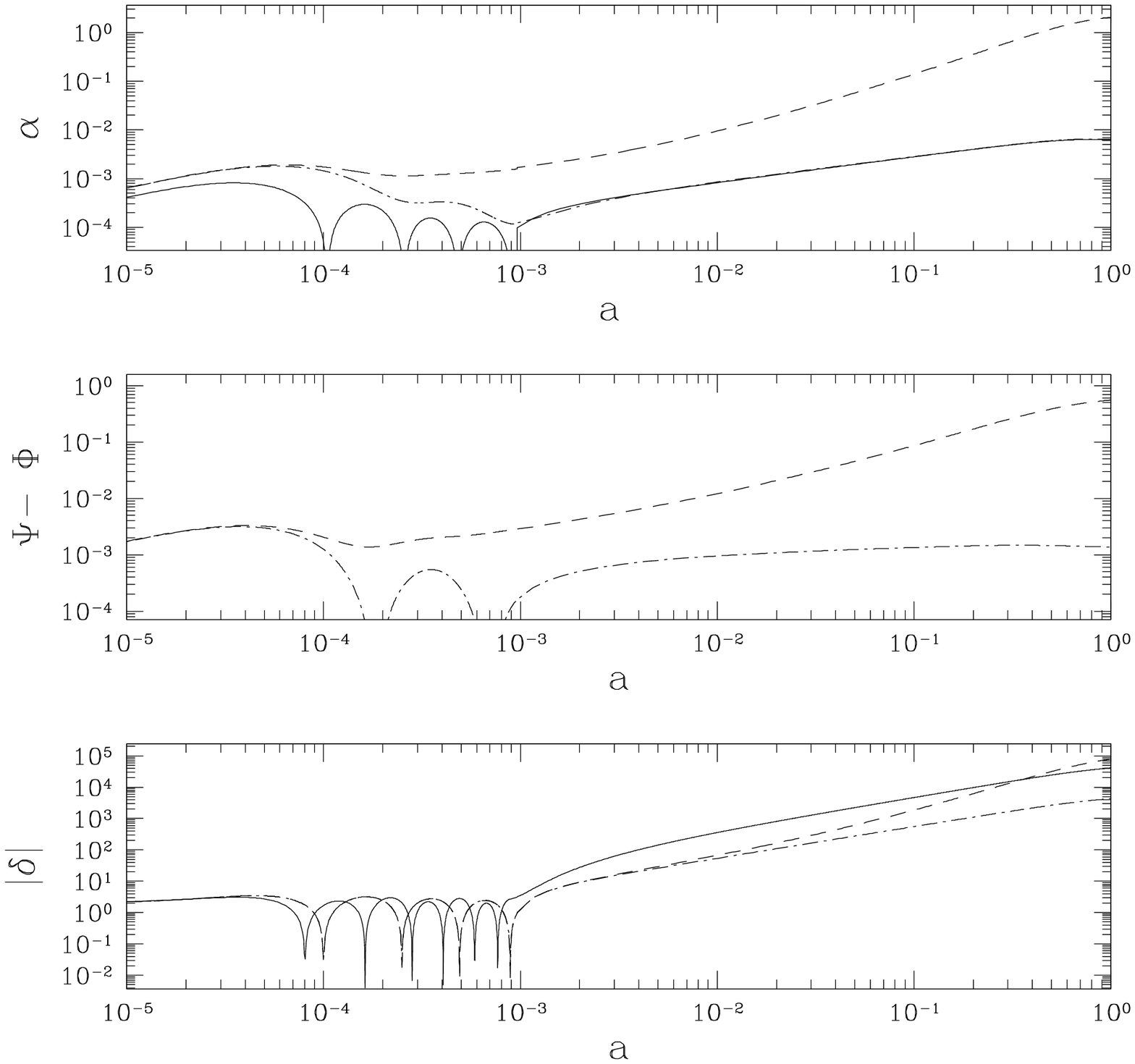}{Evolution of cosmological perturbations (unnormalized) in  
a TeVeS model with $\Omega_m=0.3$ (baryons only), $\mu_0=200$ and low value $K_B=0.07$ (dashed) and a high 
value of $K_B=1$ (dot-dashed). 
Top panel shows that vector perturbations become unstable 
for the low value of $K_B$. Solid curve is
the particular solution $\alpha=-\Psi\tau/3$. Second panel shows that this induces
a large difference between the two Newtonian potentials. Third panel 
shows that this drives enhanced growth in
the density perturbations as compared to standard $\Lambda$CDM (solid curve) if $K_B$ is small;
density perturbations in the large $K_B$ case are smaller than in $\Lambda$CDM due to
the absence of dark matter. In all cases the wavenumber is
$k=0.5$Mpc$^{-1}$}

The growing vector field drives the two Newtonian potentials to differ from one another as seen in the
middle panel of Fig.\rf{allpert}. Recall that in general relativity,
this difference is sourced only by anisotropic stress. In TeVeS, the vector field also sources the
difference~\cite{Skordis}.
This difference in turn drives 
enhanced growth in the density perturbations as shown in the bottom panel, precisely
the kind of growth needed to generate large structures from the 
small inhomogeneities present at recombination. We have verified that this growth does not occur if
vector perturbations are turned off.

How generic are these ideas of vector instabilities and their subsequent impact on
the two Newtonian potentials? 
Several authors~\cite{Lim:2004js,Zlosnik:2006zu} have pointed out that, in the context of the general
Lagrangian studied in ~\cite{Jacobson:2004ts}, vector instabilities exist for a wide range of coefficients.
So vector instability generally seems quite plausible. 
Bertschinger~\cite{Bertschinger:2006aw} has pointed out that the 
key input from modified gravity models is the source term for the
difference between the two Newtonian potentials. So it is not surprising that this difference plays an
important role in TeVeS. It is possible that the enhanced growth discussed here will emerge naturally
from a modified gravity model which explains the acceleration of the universe.

{\parindent0pt\it Conclusions.}
Motivated by observations of gravitational lensing, 
Sanders~\cite{Sanders:1996wk} was the first to introduce a vector field into the MOND framework. Bekenstein's
recent theory~\cite{Bekenstein} elevated this field to be dynamical. We have shown here that, at least in Bekenstein's framework,
the vector field can also source enhanced growth in the cosmic density perturbations. This is by no means the last word on
confronting cosmological data with MONDian theories, but it does overcome perhaps the primary cosmological hurdle 
faced by a baryon-only model. The enhanced growth enables the very small perturbations at recombination
to grow into the large structures we see today. Among the significant problems that remain are the need to match the
observed galaxy power spectrum on large scales and the well-measured series of peaks and troughs in the CMB spectrum
and the apparent mismatch between mass and light in galaxy clusters~\cite{Clowe:2006eq}.

We are very grateful to Pedro Ferreira, David Mota, and Constantinos Skordis for their help with TeVeS perturbations; Eugene
Lim for sharing his vector expertise, and Mordecai Milgrom for useful background information. 
This work was supported by the US Department of Energy and by NASA grant
NAG5-10842. ML  would like to thank the Particle Astrophysics Center for its hospitality and
the Brinson Foundation for its generous support.

\bibliography{short}

\end{document}